\documentclass{aastex}
\newcommand{\new}[1]{{{#1}}}
\usepackage{epsfig}
\begin{document}

%
   \title{On the coorbital corotation torque in a viscous disk and its
   impact on planetary migration}

   \author{F.S. Masset}

   \slugcomment{Present address: SAp, CE-Saclay, 91191 Gif/Yvette Cedex, France}

   \affil{Max-Planck-Instit\H ut f\H ur Astrophysik,
              Karl-Schwarzschild-Str, 1, Postfach 1317, 85741
              Garching, Germany\\  ({\tt fmasset@cea.fr})}

   \begin{abstract}
       We evaluate the coorbital
       corotation torque on a planet on a fixed circular orbit
       embedded in a viscous protoplanetary disk, for the case of a
       steady flow in the planet frame. This torque can be evaluated
       just from the flow properties at
       the separatrix between the librating (horseshoe) and
       circulating streamlines.
       A stationary solution is searched for the
       flow in the librating region. When used to evaluate the
       torque exerted by the circulating material \new{of the outer and inner disk} on the
       \new{trapped material} of the librating region, this solution leads to an 
       expression of the coorbital corotation torque in agreement with
        previous estimates.
       An analytical expression is given for the corotation
       torque as a function of viscosity. Lastly, we show
       that additional terms in the torque expression can play a
       crucial role. In particular, they introduce a coupling with the 
       disk density profile perturbation (the `dip' which surrounds the
       planet) and add to the corotation torque a small, positive
       fraction of the one-sided Lindblad torque. As a consequence,
       the migration could well be directed outwards in very thin disks
       (aspect ratio smaller than a few percent). This 2D analysis is
       especially relevant for mildly embedded protoplanets
       (sub-Saturn sized objects).

      \keywords{Accretion, accretion disks --
                Hydrodynamics --
                Solar system: formation -- Planetary systems: formation
                -- Planetary systems: protoplanetary disks
               }
   \end{abstract}

%

\section{Introduction}
The discovery of extra-solar planets in the last few years
renewed interest in the description of the tidal interaction of a
planet with a gaseous Keplerian disk. Indeed, many
planets, the so-called `hot Jupiters',
have been found orbiting very close to their host star, and
their formation is unlikely to have occurred at such short distances from
the star.  It is thought that these planets have formed further out
in the disk and have undergone a significant orbital decay, or
migration, as the result of the tidal interaction with the
protoplanetary disk in which they were formed. 
Goldreich \& Tremaine (1979, 1980) and Ward (1986,
1997)
have shown that most of the tidal interaction between a protoplanet
embedded in a protoplanetary disk can be accounted for by the
so-called Lindblad torques, which correspond to the exchange of
angular momentum between the planet and the disk at the position of
the Lindblad Resonances (LR) of the planet, where the \new{gravitationnal
field} of this latter excites in the disk spiral density waves which propagate
away from the planet's \new{orbit}. The Lindblad torque exerted by the
outer disk (at the Outer Lindblad Resonances, OLR) on the planet is
negative, whereas the Lindblad torque exerted by the inner disk (at
the
Inner Lindblad Resonances, ILR) on the planet is positive. Ward (1997)
has shown that the absolute value of the Outer Lindblad torque is
intrinsicly higher  than the Inner Lindblad torque; hence the net torque
(which we shall refer to hereafter  as the differential Lindblad torque) is
negative and leads to an orbital decay of the planet. 
Migration can roughly speaking be split into two regimes: 

\begin{itemize}
\item The
type~I
migration regime, in which the planet mass is small, and the disk
response is linear, that is to say that the relative perturbed density
in the planet's wake is a small fraction of unity. In this regime, the
migration velocity is proportional to the planet mass, and is
independent
of the viscosity, \new{as long as one only considers the Lindblad torque and
neglects the corotation torque} (Meyer-Vernet \& Sicardy 1987, Papaloizou \& Lin
1984). The LR pile-up \new{at high azimuthal wavenumber} at $\pm \frac 23h$ (where $h$
is the disk thickness) from the planet's orbit, and most of the 
torque acting on the planet comes from the pile-up region. The
relative mismatch between the Outer and Inner Lindblad torques, which gives
rise to migration, scales as the distance between the pile-up of the
ILR and the pile-up of the OLR, and therefore scales as the disk
aspect ratio~$h'=h/r$. Since the
one-sided Lindblad torque scales as~$h'^{-3}$ (see e.g. Ward 1997), 
the migration velocity scales as~$h'^{-2}$.

\item  When the protoplanet mass is above a certain threshold (namely
when its Hill radius is comparable to the disk thickness), 
the wake excited by the planet leads to a shock in the immediate
vicinity of the excitation region, and therefore the wake is locally
damped and gives its angular momentum to the disk.
Hence the planet can open a gap, if the viscosity is below some
threshold
value (see e.g., Papaloizou \& Lin 1984). In
that case, the disk response is markedly non-linear, and most of the
protoplanet Lindblad resonances fall in the gap and therefore
cannot contribute to the planet-disk angular momentum exchange. This
is the so-called type~II migration regime. The 
migration rate slows down dramatically compared to the type~I
migration (Ward 1997).
Furthermore, the tidal truncation splits the disk in two parts and, if
the surrounding disk mass is comparable to the planet mass, the
planet is locked in the disk viscous evolution. In this case
 the migration time
is of the order of the disk viscous timescale (Nelson et al. 2000).
\end{itemize}

One of the problems raised by type~I migration is that it is extremely
fast. In particular, the build up of a single giant planet solid core, 
which has to reach a critical mass of about
$15$~$M_\oplus$ (where $M_\oplus$ is the earth mass) before rapid gas
accretion begins (Pollack et al. 1996), is a slow process, and can be longer by more than
one order of magnitude
than its migration  timescale all the
way to the central star (Papaloizou \& Terquem, 1999). 
The planet can also exchange
angular momentum with its coorbiting material. This corresponds to the
so-called corotation torque, which is proportional to the vortensity
gradient, i.e., to the quantity $\partial (B/\Sigma)/\partial
r$,
where $\Sigma$ is the disk surface density and $B$ the second Oort's constant
(Goldreich \& Tremaine, 1979). On the
contrary of what happens at the Lindblad resonances, the flux of
angular momentum at corotation is not carried away by waves but
accumulates there. Ward (1991, 1992) has shown how the corotation
torque is due to the exchange of angular momentum between the planet
and fluid elements as they jump on the other side of the planet orbit when
they reach a hair-pin curve end of their horseshoe streamline, and that the
dependence of this torque on the vortensity gradient arises from an
unequal
mapping of surface elements between the outer and inner parts of a
horseshoe orbit. Estimates of this torque, both analytical (Ward 
1992)
and numerical (Korycanski \& Pollack 1993) show that it is at most comparable
to the differential Lindblad torque. Furthermore, in an inviscid disk,
the libration of the coorbital material tends to remove the vortensity
gradient after a time which is of the order of the turn-over time
of the outermost horseshoe orbit (which contribute most to the
coorbital corotation torque), thus switching off the corotation torque
after a transient episode. In a viscous disk, the situation is
different. \new{On one hand}, as suggested by Ward (1991), if the viscous
diffusion timescale over the horseshoe region width is smaller than
the turnover time of the horseshoe orbits, the gradient of
vortensity will not be removed. \new{On the other hand}, the material of the
outer disk, when it reaches the coorbital region during its drift
towards the center, exchanges \new{once} angular momentum with the planet as it
gets sent to the inner disk.

The purpose of this paper is to analyze such a situation. After
introducing
the problem set-up and the main notations in section~\ref{sec:not}, we
describe the streamlines in stationary regime 
(section~\ref{sec:topo}). We show that the introduction of a
finite viscosity modifies the topology of the flow around the planet,
but that there still exists a set of librating horseshoe streamlines,
which correspond to material trapped in the coorbital region, and
which therefore does not participate in the disk accretion onto the
central star. We then show in section~\ref{sec:tqform}, 
assuming that the flow is stationary in
the frame corotating with the planet, how the coorbital corotation
torque can be expressed using only the flow properties at the boundary
of the librating region (which we shall call hereafter the separatrix,
as it separates the librating from the circulating streamlines). This
simplifies greatly the torque expression and `hides' any additional
complexity arising from the flow properties interior to the
separatrix, for instance, the ``tadpole'' streamlines
around the Lagrange points $L_4$ and $L_5$.

In Appendix~\ref{app:steady}, we give a simple steady state solution for the
perturbed density in the coorbital region which neglects pressure
effects. When used to evaluate the torque expression given in 
section~\ref{sec:tqform}, we recover the expression given by Ward
(1991). We then give an analytical expression of the corotation
torque as a function of viscosity, which exhibits two different
regimes,
depending on the ratio of the turnover to viscous timescales of the
horseshoe region. A third regime is also mentioned, which corresponds
to a cut-off at high viscosity when some of the streamlines
originating
from the outer disk arrive at the inner disk without having passed by
the planet, i.e.,  when the viscous drift time over the horseshoe region
is shorter than the horseshoe turnover time.

In section~\ref{sec:coupl}, we show that the full corotation torque
expression as given in section~\ref{sec:tqform} exhibits
additional terms, which have potentially important effects. Indeed,
they correspond to a small positive fraction of the one-sided Lindblad
torque, and this fraction turns out to be independent both of
viscosity and disk thickness. Since the relative mismatch between
Inner and Outer Lindblad torques scales as $h'$, this additional term can
be larger in absolute value than the 
differential Lindblad torque in sufficiently
thin disks. This term, which can be estimated only in order of magnitude,
may therefore significantly slow down, or possibly reverse, type~I migration in
very thin disks.


\section{Problem set-up and notations}

\label{sec:not}

We consider a non-accreting protoplanet on a fixed circular orbit of
radius $r_p$ in a disk with uniform and constant
 aspect ratio $h'=h/r$ and viscosity~$\nu$.
Since the planet is considered as non-accreting, its mass~$M_p$ has to be
below a threshold of about $15$~$M_\oplus$ (Papaloizou \& Terquem, 1999).
 The disk surface density
is denoted $\Sigma$,  and the
unperturbed disk surface density is uniform and is denoted
$\Sigma_\infty$. It is also the value far from the planet. The
viscosity is also given by:
$\nu=\alpha \Omega h^2$, where $\Omega$ is the disk orbital
frequency, and $\alpha$ is a dimensionless parameter (Shakura \& Sunyaev 1973).
The planet orbital frequency is denoted $\Omega_p$. A fluid element is
labeled in the disk by its coordinates $r$ and $\theta$
in a frame corotating with the planet, where $r$ is the distance to
the primary and $\theta$ the azimuth (the planet is at azimuth~$0$),
counted rotation-wise. The perturbed azimuthal velocity in the
presence of the planet is denoted $v$, and
$j=r^2\Omega+rv$ is
 the specific angular
momentum of a fluid element. Its value in the
unperturbed disk is  $j_0=r^2\Omega$.
\new{We shall also make an extensive use of the distance to the orbit
$x=r-r_p$, and its non-dimensional counterpart $\hat x=x/r_p$.}


\section{Streamline topology}

\label{sec:topo}
\subsection{Inviscid case}

\begin{figure}
\psfig{file=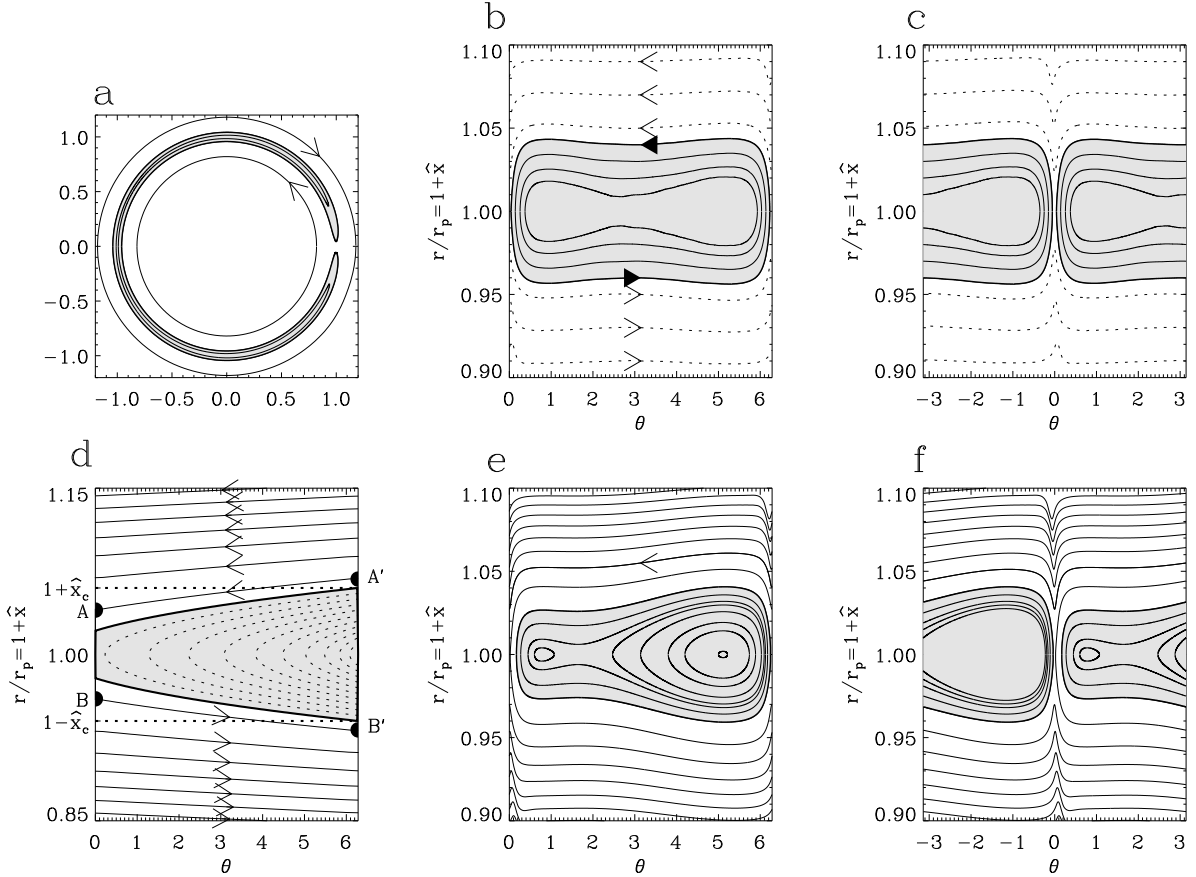,width=\columnwidth}
\caption{\label{fig:nouv_fig}The figures $a$ to $c$ represent the same
situation: a $15$~M$_\oplus$ protoplanet embedded in a $4$~\%
aspect ratio disk, once a steady state has been reached. The details
of the run are unimportant here, as these figures are for illustration
purpose only, and closely resemble other runs presented elsewhere (see
e.g. Masset \& Snellgrove 2001, or Nelson et al. 2000). Figure~$d$ shows
streamlines which fulfill the prescriptions~1 to~3 of section~
\ref{sec:topology2}. The figures~$e$ and~$f$ represent the same situation,
except for a $\pi$-shift in azimuth. The run they are extracted from differs
from the run of figures~$a$ to~$c$ only by the viscosity, which has been set
this time to a very high value. In figures~$d$ to~$f$, the solid line in
the outer and inner disk (outside of the shaded --librating-- region)
represents a unique streamline, which is also, as we are dealing with
a steady flow, the path of a fluid element on its way to the central star.}
\end{figure}

In an inviscid disk, the streamlines in the coorbital region of a
planet on a fixed circular orbit
have the aspect represented in Fig.~\ref{fig:nouv_fig}a,
\ref{fig:nouv_fig}b and~\ref{fig:nouv_fig}c. The
shaded
region represents the horseshoe streamlines. 
The streamlines in the outer and
inner disk, which are represented as dotted lines
in Fig.~\ref{fig:nouv_fig}b and~\ref{fig:nouv_fig}c, are circulating,
and closed in the case of a steady flow (as can
be seen in Fig.~\ref{fig:nouv_fig}a, where these streamlines
appear as almost circular streamlines), 
since there is no radial drift (Lubow, 1990).
The separatrix between the librating (horseshoe) and circulating
(inner and outer disk) streamlines is represented as a thick solid
line. The planet lies at $r/r_p=1$ and $\theta=0$ (mod. $2\pi$).
The circulating streamlines always stay on the same side of the
planet's orbit, as is better seen in Fig.~\ref{fig:nouv_fig}c.
One can see that these circulating streamlines get deflected as they
pass by the planet. This corresponds to the excitation of epicyclic
motion and the exchange of angular momentum with the disk, which is
carried away by pressure effects and gives rise to the wake, which can
be clearly seen in Fig.~\ref{fig:nouv_fig}b and~\ref{fig:nouv_fig}c.
The angular momentum transfer rate into this
wake corresponds to the Lindblad torques.

\subsection{Viscous case}
\label{sec:topology2}

\new{In order to have a first idea of} the streamline topology
we  approximate the action of the planet on
the flow with the following prescription:
\begin{enumerate}
\item The planet acts on a fluid element at $\theta=0$ (mod. $2\pi$) only (i.e., at a conjunction).
\item If the distance~$|x|$ of the fluid element to the orbit when
it is in conjunction with the planet, is
smaller than some threshold value $x_c$ (which corresponds to the
half-width of the horseshoe region), then the fluid element is
``reflected'' with respect to the planet orbit and sent to 
$(\theta, -x)$. Otherwise no action is taken on the fluid element.
\new{This is meant to mimic the U-turn of the fluid element at the end of the
horsestreamlines in the horseshoe region.}
\item The fluid element velocity is assumed to be everywhere the
velocity in the unperturbed disk.
\end{enumerate}

Obviously this \new{over}simple prescription reproduces in an inviscid case
the main properties
outlined at the previous section, i.e., it gives rectangular librating
streamlines in the horseshoe region \new{(the width of which is $2x_c$)}, and circulating circular
streamlines in the outer and inner disk. However there is no wake.
In a viscous disk \new{with uniform viscosity and surface density}, 
the fluid element velocity in the corotating frame
is given by (if we assume the radial extend of the coorbital region to
be small compared to the planet distance):
\begin{eqnarray}
\label{eqn:sketch_ez}
\dot x = -\frac 32 \frac \nu{r_p}\mbox{~~~and~~~}
\dot\theta = -\frac32\frac{\Omega_p}{r_p}x
\end{eqnarray}
which integrates as: 
\begin{equation}
\label{eqn:simplsl}
\theta=\frac
12\frac{\Omega_p}{\nu}x^2+\theta_0
\end{equation}
The streamlines are therefore arcs of parabola, and the sketch of the
streamlines
which respect the prescription given above is represented in
Fig.~\ref{fig:nouv_fig}d.
If we follow the path of a fluid element along the solid streamline
originating in the outer disk, we see that first this fluid element
has a circulating trajectory and approaches progressively the planet
coorbital zone. The distance between two successive intersections of
the streamline with say the $\theta=0$ axis increases steadily,
since the angular frequency mismatch between the planet and the fluid
element decreases. When the fluid element reaches the point~$A'$, it
is not sent on the other side of the orbit, since its distance to the
planet is still larger than $x_c$. However, along the arc $A'A$, it
crosses the horseshoe region outer boundary, and when it reaches~$A$, it
is sent to~$B$ and therefore gives to the planet, during this close encounter, the
positive amount of angular momentum that it loses. It then follows its
path along $BB'$, which by construction is symmetric to $AA'$, 
therefore when it reaches $B'$, it is out of the horseshoe region,
\new{and therefore it} keeps
circulating in the inner disk, and eventually gets accreted onto the
primary.

One can wonder how this schematic picture changes in a more realistic
case. First it can be useful to notice that in a 2D steady flow
defined on a compact domain, a streamline which does not intersect the
domain boundaries is closed\footnote{One reason for that is that the
field
$\Sigma \vec u$, where $\vec u$ is the velocity field, is divergence
free, and can therefore be expressed as the curl of a field
$\psi\vec{e_z}$,
where $\psi$ is a scalar field and $\vec {e_z}$ a vector orthogonal to
the disk plane. \new{Hence} the function $\psi$ is conserved along the field lines of
$\Sigma \vec u$, which are also the streamlines. The streamlines
therefore appear as isocontours of the function~$\psi$.}. Therefore
\new{one has to expect} a region of closed streamlines behind the planet, which
enclose the fixed point on the orbit where the negative viscous torque
cancels out the positive torque from the planet, which goes from $0$
to an arbitrarily large value as $\theta$ goes from~$\pi$ to~$2\pi$,
and,
if the viscosity is small enough, this closed region \new{should} extend from
one end to another of the horseshoe region. We represent in 
Fig.~\ref{fig:nouv_fig}e and~\ref{fig:nouv_fig}f a situation 
taken from a run similar to
the run
of Fig.~\ref{fig:nouv_fig}a to~\ref{fig:nouv_fig}c, except that
the viscosity has been set to a very high value (corresponding to
$\alpha\sim 0.1$, which is probably too high for a
protoplanetary disk, but has the merit that it clearly shows the
streamlines topology).
We see on this figure that we have a closed streamlines region which resembles the one
obtained with the \new{oversimple} prescription, and that material
flows from the outer disk to the inner disk.

Therefore, in the case of a finite viscosity,  we still have a set of closed
librating streamlines in the coorbital region, enclosed by a closed
separatrix,
which corresponds to trapped material which will not be
accreted onto the primary, but now the outer and inner disk
communicate, and as material flows from the outer disk to the inner
disk it exerts on the planet a positive torque. The smaller
the viscosity, the more the librating region  has a rectangular
shape
and resembles the horseshoe region of the inviscid case.

\section{Coorbital corotation torque expression}

\label{sec:tqform}
The material enclosed within the separatrix is trapped and therefore
its angular momentum is constant with time, since the planet is on a
fixed circular orbit. Hence the net torque on
the librating region~$L$ vanishes. This torque can be decomposed as the
gravitational torque exerted by the planet~$\Gamma_{p\rightarrow L}$
and the viscous torque~$\Gamma^{vis}_S$ exerted by the
inner and outer disk on~$L$ at the separatrix~$S$, hence:
\begin{equation}
\label{eqn:vanish}
\Gamma_{p\rightarrow L}+\Gamma^{vis}_S=0
\end{equation}
In what follows we make
the following assumptions:
\begin{enumerate}
\item the viscosity is small enough so that the
separatrix distance to the orbit is almost constant when one stands
far from the horseshoe U-turn regions;
\item the
outer and inner separatrices lie at the same distance $x_s$ from the orbit.
\new{This neglects in particular the mismatch between the orbit and the corotation,
due to the fact that the disk is slightly sub-Keplerian (due to the partial support of
the central object gravity by a radial pressure gradient). For an aspect ratio
of a few percents, this mismatch is typically $10^{-3}r_p$, while for the protoplanet
masses we consider, $x_s \sim $~a few~$10^{-2}r_p$, and therefore assumption~2
is reasonable. However, when considering disks with aspect ratios $> 0.1$, and/or
smaller mass planets, this mismatch can no longer be neglected.}

\item the perturbed quantities in the librating
region are independent of $\theta$, and thus are only function of
$x$. 
\end{enumerate}
Experience from numerical
simulations shows that this last assumption is reasonable; for instance,
the shallow `dip' profile that is observed when the mass is sufficiently large
and the viscosity is sufficiently low can be considered as
independent of $\theta$ with a very good approximation as long as one
excludes the U-turn ends of the horseshoe region.
We neglect the torque exerted by the inner and outer disk
on~$L$ at the U-turn ends of the librating region, \new{and we neglect the torque
of the pressure forces on~$L$}. The coorbital corotation
torque on the planet, which is the torque exerted by the disk material
during the close encounters with the planet\footnote{\new{A point of terminology
is needed here: in an inviscid situation, the coorbital corotation
torque simply
corresponds to the exchange of angular momentum with the fluid elements
which corotate, in average, with the planet, i.e. material strictly located
at the corotation in the linear limit, and, in the case of a finite mass,
the whole material of the horseshoe region (indeed, as a fluid element
located in this region can never be in conjunction with the planet, when
this latter has described $N$~orbits in an inertial frame, the fluid
element has described at least $N-1$~orbits, and at most $N+1$~orbits; 
taking $N\rightarrow\infty$, one is led to $\langle\Omega\rangle=\Omega_p$, where
$\langle\Omega\rangle$ is the average angular velocity of the fluid element). In a
viscous case, the positive torque arising from the close encounters of
the outer disk fluid elements with the planet as they get sent to 
the inner disk needs to be taken into account, even if these fluid 
elements are only temporarily corotating with the planet. Failing to do so
would in particular make it impossible to reconcile the torque expression
given in this paper with previous estimates, since this positive torque
almost cancels out the negative viscous torque exerted on the librating
region; the net torque, which we identify as the coorbital corotation torque,
comes from a balance between these two.
}}, can therefore be
expressed, using Eq.~(\ref{eqn:vanish}) as:
\begin{eqnarray}
\label{eqn:ct1}
\Gamma_C&=&\Gamma_{L\rightarrow p}+\Gamma_R=\Gamma^{vis}_S+\Gamma_R{}\\
\nonumber
&=&{}\Gamma^{vis}_{S^+}+\Gamma^{vis}_{S^-}+\Gamma_R
\end{eqnarray}
where $\Gamma_{L\rightarrow p}=-\Gamma_{p\rightarrow L}$ is the
gravitational torque exerted by the librating region on the planet,
$\Gamma_R$ is the torque exerted \new{on the planet} by the circulating material when it
flows from the outer disk to the inner disk, $\Gamma^{vis}_{S^+}$ is
the torque exerted by the outer disk on the librating region at 
the outer separatrix, and
$\Gamma^{vis}_{S^-}$ is the torque exerted by the inner disk on the
librating
region at the 
inner separatrix. We can write:
\begin{equation}
\label{eqn:vtp}
\Gamma^{vis}_{S^\pm}=\mp 2\pi\nu(r_p\pm x_s)^2\Sigma_{S^\pm}\left[-\frac 32
\Omega
+\left(\frac{\partial v}{\partial r}\right)\right]_{r_p\pm x_s}
\end{equation}
and:
\begin{equation}
\label{eqn:gammar1}
\Gamma_R=\dot M[j(r_p+x_s)-j(r_p-x_s)]
\end{equation}
where $\dot M$ is the mass flow from the outer disk to the inner disk,
which would be $3\pi\nu\Sigma_\infty$ in a \new{strictly} 
stationary regime, and
where we have assumed: $|\partial v/\partial r| \gg |v/r|$.
To
lowest order in $x_s/r_p$, Eq.~(\ref{eqn:gammar1}) can be recast as:
\begin{equation}
\label{eqn:gammar2}
\Gamma_R=\dot M[\Omega_pr_px_s+2(r_p+x_s)v_+-2(r_p-x_s)v_-]
\end{equation}
where $v_\pm$ is the perturbed azimuthal velocity at the outer
(resp. inner) separatrix.
Therefore, the use of Eqs.~(\ref{eqn:ct1}), (\ref{eqn:vtp}) and 
(\ref{eqn:gammar2}) enables one to estimate the corotation torque only
from the knowledge of the unperturbed disk characteristics and the
flow properties at the separatrix, provided a steady flow solution is
known, since the steady flow assumption is necessary to 
write~Eq.~(\ref{eqn:vanish}).

\section{Steady state solution}

\label{sec:steady}

\label{sec:libration}
In order to find a steady state solution, we write a radial \new{viscous} iffusion
equation for the surface density. This equation needs 
 to include the turnover of the horseshoe
orbits. One simple way to do that is to include source and sink terms
for $|x|<x_s$. This is detailed in Appendix~\ref{app:steady}.

When this steady flow solution is used to evaluate the torque
expression
given by Eqs.~(\ref{eqn:ct1}), (\ref{eqn:vtp}) et (\ref{eqn:gammar2}),
in which we neglect the perturbed azimuthal velocity profile and the
even part in $x$ of the density profile perturbation (the `dip' which
surrounds
the \new{orbit}), one obtains:
\begin{equation}
\label{eqn:gammac}
\Gamma_C=\frac 92 x_s^4\Omega_p^2\Sigma_\infty {\cal F}(z_s)
\end{equation}
where:
\begin{equation}
\label{eqn:deff}
{\cal F}(z_s) = \frac{1}{z_s^3}-\frac{g(z_s)}{z_s^4g'(z_s)}
\end{equation}
where $g$ is \new{a linear combination of the Airy functions~Ai and~Bi} defined by Eq.~(\ref{eqn:defg}) and where:

\begin{equation}
z_s=x_s\left(\frac{\Omega_p}{2\pi\nu r_p}\right)^{1/3}
\end{equation}

In particular, in the highly viscous case ($z_s\ll 1$),
we have:
\begin{equation}
\label{eqn:hv}
\Gamma_C=\frac 98 x_s^4\Omega_p^2\Sigma_\infty
\end{equation}
where we use the fact that (Abramowitz \& Stegun, 1972): 
\begin{equation}
\frac{g(t)}{g'(t)}=t-\frac 14 t^4+o(t^4)
\end{equation}
Eq.~(\ref{eqn:hv}), with our notations, is the same as the expression
given by Ward (1992):
\begin{equation}
\label{eqn:torq_ward}
\Gamma_C'=\frac 34x_s^4\Omega_p^2\Sigma_\infty\left(\frac
32-\beta\right)
\end{equation}
in which we have to set the slope $\beta = -d\log\Sigma/d\log r$ to $0$ since we
are in the very viscous regime in a uniform surface density disk.

\begin{figure}
\psfig{file=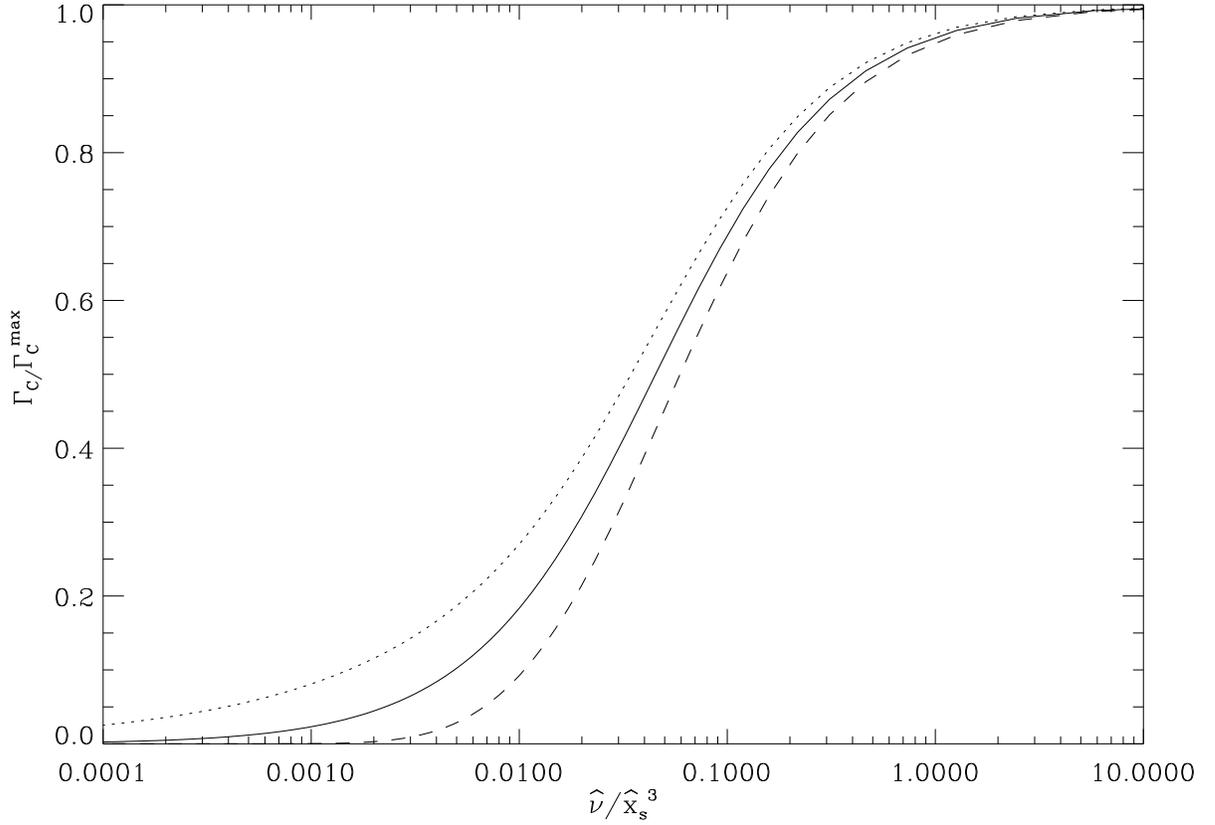,width=\columnwidth}
\caption{\label{fig:gamma}Coorbital corotation torque as a function of
viscosity. The solid curve is a plot of expression~(\ref{eqn:gammac}),
while the dashed and dotted ones represent respectively
Eqs.~(\ref{eqn:torq_ward_1})
and~(\ref{eqn:torq_ward_2}). Here $\Gamma_C^{max}$ stands for
$\frac98\Omega_p^2x_s^4\Sigma_\infty$.}
\end{figure}

In the general case, at lower viscosities, there is an ambiguity in
the choice of the slope of $\beta$ that has to be used in
Eq.~(\ref{eqn:torq_ward}). Indeed, if we choose:
\begin{equation}
\label{eqn:beta1}
 \beta_1=-\frac{d\Sigma}{dx}\frac{r_p}{\Sigma_\infty}
\mbox{~~~at
$\hat x=0$}
\end{equation}
then we get the expression:
\begin{equation}
\label{eqn:torq_ward_1}
\Gamma_{C_1}'=\frac
98x_s^4\Omega_p^2\Sigma_\infty\frac{g(z_s)}{z_sg'(z_s)}
\end{equation}
whereas if we choose:
\begin{equation}
\label{eqn:beta2}
 \beta_2=-\frac{\Sigma(x_s)-\Sigma(-x_s)}{2x_s}\frac{r_p}{\Sigma_\infty}
\end{equation}
then we get:
\begin{equation}
\label{eqn:torq_ward_2}
\Gamma_{C_2}'=\frac
98x_s^4\Omega_p^2\Sigma_\infty\frac{g'(0)}{g'(z_s)}
\end{equation}
The graphs of $\Gamma_C$, $\Gamma_{C_1}'$ and $\Gamma_{C_2}'$ are
represented in Fig.~\ref{fig:gamma}. $\Gamma_{C_1}'$
is always smaller than $\Gamma_C$, and  $\Gamma_{C_2}'$ is always
larger than $\Gamma_C$. This means that it is always possible to find
a value for $\beta$ between $\beta_1$ 
and  $\beta_2$ for which
Eqs.~(\ref{eqn:gammac})
and~(\ref{eqn:torq_ward}) give the same value.
The ambiguity in the choice of $\beta$ for Eq.~(\ref{eqn:torq_ward})
comes on one hand from the fact that this equation has been derived by an
integration of the torque on all the librating streamlines of the
horseshoe region, and on the other hand from the fact that the
solution of Eq.~(\ref{eqn:sol1})
does not lead to a constant slope over the whole horseshoe region. 
Such an ambiguity does not exist in Eq.~(\ref{eqn:gammac}) since this
latter has been derived taking into account only the disk perturbation
at the separatrix. What happens in the interior of the librating
region does not need to be known, provided we are dealing with a steady flow.

The ratio ${\cal R}=\nu r_p/(\Omega_p x_s^3)=1/(2\pi z_s^3)$ is also
$\frac{1}{8\pi}\frac{\tau_{HS}}{\tau_{visc}}$,
where:
\begin{equation}
\label{eqn:tauHS}
\tau_{HS}=\frac{4\pi r_p}{\frac 32\Omega_p x_s}
\end{equation}
and:
\begin{equation}
\label{eqn:tauvisc}
\tau_{visc}=\frac{x_s^2}{3\nu}
\end{equation}
are respectively the turnover time of the outermost horseshoe orbit and the
viscous diffusion time across the horseshoe region. From
Fig.~\ref{fig:gamma}
we see that the coorbital corotation torque reaches half of its
maximum value at ${\cal R} = \hat\nu/\hat x_s^3\sim 0.05$, that is to say
for $\tau_{HS}\sim\tau_{visc}$. When the turnover time is much larger
than the viscous time, i.e. when ${\cal R} \gg 0.05$, the
corotation torque reaches its maximal value, since libration is
inefficient at removing the vortensity gradient. On the contrary, when
the viscous time is much larger than the turnover time (i.e.,  when
${\cal R}
\ll 0.05$) libration can remove the vortensity gradient and the
corotation torque has a negligible value. 

\subsection{Cut-off at high viscosity}

As discussed in section~\ref{sec:topology2}, the outer separatrix can be
approximated
as a branch of parabola in the $\theta,r$ plane, and so far we have
assumed that the viscosity is small enough so that we can
approximate the separatrix by a line of constant distance to the
orbit. However, if the viscosity is very high this is no longer true, and
the separatrix can reach the
orbit before having described a complete revolution in the planet
frame.
From Eq.~(\ref{eqn:simplsl}), we can give an estimate of the critical value
for the 
viscosity above which this will happen, by setting $\theta_0=0$,
$x=x_s$ and $\theta=2\pi$. This leads to:
\begin{equation}
\label{eqn:critvisc}
\nu_c=\frac{x_s^2\Omega_p}{4\pi}
\end{equation}
The 2D analysis we have developed here is mainly valid for
mildly embedded protoplanets, for which the Hill radius is a sizable
amount of the disk thickness; this would correspond to values of the
$\alpha$-parameter of about $\alpha\sim 0.1$, probably too
high for protoplanetary disks, and therefore this regime is not very 
relevant for the analysis described in this paper.

\section{Additional terms and Lindblad torque coupling}
\label{sec:coupl}
The coorbital corotation torque general expression in a steady regime,
which we have given in Eqs.~(\ref{eqn:ct1}), (\ref{eqn:vtp}) and
(\ref{eqn:gammar2}), contains additional terms that we have so far neglected.
In the following we still neglect the terms linked to the
perturbed azimuthal velocity~$v$, the analysis of which will be found
in Appendix~\ref{otherterms}, but we do not neglect, as we have
done before, the perturbation of the density profile in the vicinity
of the planet orbit. Indeed, as is seen frequently in numerical
simulations, the mildly embedded planets, which are not massive enough
to fulfill the gap opening criteria, are nevertheless surrounded by a shallow
axisymmetric depletion of their coorbital region. As a consequence, a
new term arises in the torque expression, which reads:
\begin{equation}
\label{eqn:comp1}
\Gamma_C^{I}=3\pi\nu(\Sigma_\infty-\Sigma_s)\cdot\Omega_pr_px_s
\end{equation}
where $\Sigma_s$ is the average of the surface density at the outer
and inner separatrices (in different words it is the value of the even
part in $\hat x$ of the perturbed density, sampled at any separatrix).
If we assume that this value is the same as the value at the orbit
(which means that the dip edges lie further from the orbit than the
separatrix),
then we can write:
\begin{equation}
\label{eqn:densjump}
\Sigma_\infty-\Sigma_S=\frac{\Gamma_{LR}}{3\pi\nu\Omega_pr_p^2}
\end{equation}
where $\Gamma_{LR}>0$ is the one-sided Lindblad torque. Eq.~(\ref{eqn:densjump})
is obtained by equating the viscous torque and the torque given
by the wake damping, and integrating it between the orbit and an outer
radius where most of the wave angular momentum has been transfered to
the disk (\new{i.e., it is obtained by integrating twice the even
part in $x$ of Eq.~(\ref{eqn:diff1}) in the steady state case}). 
Because of this integration, the assumption of local damping
is not necessary in writing Eq.~(\ref{eqn:densjump}). However, the
damping must occur over a radial range smaller than the planet
orbital radius \new{for this expression to be accurate}.
Recent work by Goodman \& Rafikov (2000) shows
that this may be indeed the case even for relatively small planet
masses.
Using Eq.~(\ref{eqn:densjump}), the torque expression of
Eq.~(\ref{eqn:comp1})
can be recast as:
\begin{equation}
\label{eqn:comp1_bis}
\Gamma_C^{I}=\frac{x_s}{r_p}\Gamma_{LR}
\end{equation}
It can be argued that this result has been obtained assuming a
strictly steady flow, and in particular that the mass flow \new{from the outer
disk to the inner disk} across the
planet orbit is exactly $3\pi\nu\Sigma_\infty$. \new{A higher mass rate than
this value (which is $3\pi\nu\Sigma_\infty$ either far from the planet 
or in the unperturbed
disk) would lead to a value for 
$\Gamma_C^{I}$ higher than the one
given by Eq.~(\ref{eqn:comp1_bis}). The analytical work for an inviscid disk
by Lubow (1990),
and Goodman \& Ravikov (2000), as well as numerical evaluation of the
mass flow rate accross the orbit of a non-accreting protoplanet (Masset
and Snellgrove, 2001, in their Fig.~4) suggest that the actual
mass flow rate across corotation is higher than $3\pi\nu\Sigma_\infty$,
and that the difference between the actual value and $3\pi\nu\Sigma_\infty$
scales with $\Sigma_\infty M_p^2$. This issue
definitely requires further work, as it could provide another positive
additional term to the corotation torque.}

The other terms which appear in the general torque expression are
evaluated in~\ref{otherterms}. They all correspond to the difference
between the outer and the inner separatrix of
quantities
involving the perturbed azimuthal velocity~$v$. The perturbed
azimuthal velocity can be split in two parts: the part
which corresponds to the disk axisymmetric profile perturbation under
the action of the Lindblad torques, which is odd in~$x$ to lowest
order in the disk aspect ratio, and an even part in $x$ (to lowest
order in $\hat x_s$), which corresponds to the odd perturbation density
due to libration that we have analyzed at
Appendix~\ref{app:steady}. However in the analysis of 
Appendix~\ref{app:steady} 
we have
neglected the perturbation of the azimuthal velocity, and therefore we
cannot estimate the additional terms linked to the even component of
the perturbed azimuthal velocity. Taking this latter into account would have
led to a higher order diffusion equation in Eq.~(\ref{eqn:diffset})
which would have involved the disk thickness, which has not been
considered in the simple analysis of Appendix~\ref{sec:libration}. This
will be done in a forthcoming work. We note that the
additional term linked to this even component of the perturbed 
azimuthal velocity
may play a non negligible role and significantly modify the viscosity
dependent part of the corotation torque
given by Eq.~(\ref{eqn:gammac}).

The additional term linked to the odd component of the perturbed
azimuthal velocity depends only on the value of this latter at the
separatrix. 
 However, one can only give a rough estimate of the
maximal value of these terms. Their actual value will depend on the
exact position at which the separatrix samples the perturbed velocity
profile.
It turns out that the maximal value of this term is of the same
order of magnitude as the additional term of
Eq.~(\ref{eqn:comp1_bis}):
\begin{equation}
\label{eqn:comp2}
\Gamma_{C}^{II}\le\frac 23\frac{x_s}{r_p}(2-\beta)\Gamma_{LR}
\end{equation}
Hence we see that the main effect of the additional terms (both
$\Gamma_C^I$
and $\Gamma_C^{II}$) is to
introduce a coupling with the one-sided Lindblad torque, as
they add a small, positive
fraction of this latter to the coorbital corotation torque.
Now, the differential Lindblad torque is a small, negative fraction of
the one-sided Lindblad torque, which scales with the aspect ratio (for
instance in the disk we consider, with a uniform surface density and a
constant aspect ratio~$h'$, $\Delta\Gamma_{LR}\approx
-8h'\Gamma_{LR}$).
Therefore this additional term plays against the differential
Lindblad torque and 
%
one has to expect that the migration
is considerably slower than standard type~I migration
estimates, or even reversed, over a certain planet mass range, in very
thin viscous disks. The exact determination of this mass range has to
be achieved through numerical simulations, since the analytical
determination of the exact perturbed density profile, 
of the exact azimuthal perturbed
velocity profile, and of the exact separatrix position seems a hardly
tractable task.

\section{Discussion}

We have derived the expression of the coorbital corotation torque
exerted by a uniform viscosity and surface density disk on an
embedded protoplanet on a fixed circular orbit, assuming a 2D steady flow
in the planet frame. We have obtained \new{the following} 
expression for the torque:

\begin{equation}
\label{eqn:finale}
\Gamma_C=\frac 92x_s^4\Omega_p^2\Sigma_\infty{\cal F}(z_s)+\frac{x_s}{r_p}
{\cal G}(x_s)\Gamma_{LR}
\end{equation}

\new{where ${\cal F}(z_s)$ is given by Eq.~(\ref{eqn:deff}), and where 
${\cal G}(x_s)$ is a function of the order of unity, the exact value
of which depends on the position of the separatrix and of the perturbed
axisymmetric profile of the disk under the action of the Lindblad torques,
and the maximum value of which is $7/3$, as can be inferred from 
Eqs.~(\ref{eqn:comp1_bis}) and~(\ref{eqn:comp2}). The first term in the
R.H.S. of Eq.~(\ref{eqn:finale}) corresponds to previous estimates of the
corotation torque, and exhibits an explicit dependence on the viscosity:
when the viscous diffusion timescale across the coorbital region is larger
than the horseshoe libration time, this term is negligible, while when
the viscous diffusion time is shorter than the libration time, it reaches
its maximum value. The second term of the R.H.S. of Eq.~(\ref{eqn:finale})
comes from a dependence of the viscous torque acting on the boundary (separatrix)
of the trapped librating region on the perturbed axisymmetric profile of
the disk under the action of the one-sided Lindblad torques on each side of 
the orbit.}
This term slows down the migration and could even reverse it
in sufficiently thin disks.

This analysis is valid as long as the flow around the planet can be
considered
as a 2D steady flow. Therefore the planet Hill radius needs to be a
sizable amount of the disk thickness, and gas accretion cannot occur
onto the planet (otherwise the flow cannot be considered as steady if
the accretion time is short, and the separatrix splits into two lines which
enclose the band of the material which is going to be accreted; see
e.g.,
Lubow et al. 1999). Therefore the mass limit for this analysis is roughly
$15$~$M_\oplus$. Another condition for the analysis presented above to
be valid is that the planet does not open a gap in the disk (otherwise
the corotation torque is negligible). The depletion which appears
around the planet orbit and which leads to the Lindblad torque coupling
needs to be shallow.

The analysis presented here relies on the existence of a viscous drift
of the disk material across the planet orbit. Therefore it can
apply to those parts of the disk where MHD turbulence occurs and leads
to an effective viscosity (Balbus \& Hawley, 1991). 
Hence it is likely to apply mainly to the
innermost part of the protoplanetary disks (the central astronomical
unit). \new{It is questionable whether this kind of turbulent transport
can be correctly modeled by an effective viscosity when applied to the
disk-planet angular momentum exchange in the coorbital region. 
In particular the steady state assumption that was used to estimate
the corotation torque needs to be reconsidered.}

One may speculate about the extrapolation of the calculations exposed
here to smaller masses, for which the Hill radius is much smaller than
the disk thickness (deeply embedded planets). However the
2D flow assumption was essential for the identification of a closed
region
of trapped material. There is no guarantee that we \new{could perform the same
kind of calculation in the 3D case. However, if the planet has no inclination
and no eccentricity, the gas motion in the disk equatorial plane would still
be 2D for symmetry reasons, and therefore would have a set of closed
streamlines in the steady state regime. However, the extension of the trapped
region above the equatorial plane would need to be discussed.}

\new{In the case of a finite excentricity, the analysis exposed here is no longer valid;
in particular, the planet and the disk can exchange angular momentum at the
excentric corotation resonance locations, which are not localized where the disk
material corotates with the planet guiding center (i.e., which do not lie at the orbit 
semi-major axis, if one
neglects the radial pressure gradient as a partial support to gravity in the disk). 
However, it would be interesting to find a link between the torque expression
obtained by summing over the resonances and an analysis similar to the one 
exposed above, in which the separatrix would become a band of width $2ea$ (using
 the fact that the epicyclic frequency is much
higher than the outermost horseshoe turnover frequency). In particular it would be
of interest to see how the coupling with the one-sided Lindblad torque is modified
by the introduction of a finite excentricity.}

This analysis needs to be extended to the case of an arbitrary surface
density profile, and to a migrating planet (for which the mass flow
across the orbit will therefore differ from
$3\pi\nu\Sigma_\infty$). In this case the corotation torque would include
a \new{delayed} term proportional to $\dot a$, the planet drift rate, and would
lead to a feedback on the migration. This will be presented in a
forthcoming work.

\appendix
\section{Steady flow solution}
\label{app:steady}

In the horseshoe region we write the continuity equation integrated over an arc
of a ring $[r,r+\delta r]$ which excludes the horseshoe U-turn ends of the streamlines.
We are led to:
\begin{equation}
\label{eqn:conti001}
\frac{\partial\Sigma}{\partial t}+\frac{1}{r}\frac{\partial(\Sigma ur)}{\partial r}-
\frac{1}{2\pi}\left(\frac{r'}{r}|\Omega_p
-\Omega'|\Sigma'
\frac{\delta r'}{\delta r}
-|\Omega_p-\Omega|\Sigma\right)=0
\end{equation}
where $u$ is the radial velocity. In Eq.~(\ref{eqn:conti001}) we assume that the arc of ring
spans almost an angle $2\pi$, and that it is mapped, after a horseshoe U-turn, onto the
arc of ring of radius $r'$ and width $\delta r'$. We also neglect the perturbed azimuthal 
velocity.
The variables $\Sigma'$ and $\Omega'$ stand respectively
for $\Sigma(r')$ and $\Omega(r')$. The left term of the
bracket corresponds to the mass which comes from `the other
side' (i.e.,  the ring $[r',r'+\delta r']$), and enters the ring $[r,r+\delta r]$ 
after a horseshoe U-turn,  while
the second term in this bracket corresponds to the mass which
leaves the ring $[r,r+\delta r]$.

The torque per unit mass $T$ acting on the disk material can be decomposed as the sum of the
viscous torque per unit mass and the torque per unit mass  $(2\pi R\Sigma)^{-1} 
d\Gamma/dr$ arising from the
wake damping:
\begin{equation}
\label{eqn:tqunitmass}
T=\frac{1}{r\Sigma}\frac{\partial(r^3\nu\Sigma\frac{d\Omega}{dr})}{\partial r}
+\frac{1}{2\pi R\Sigma}\frac{d\Gamma}{dr}
\end{equation}
Writing $T=u\partial j/\partial r=u\partial j_0/\partial =ur\Omega/2$, and using Eqs.~(\ref{eqn:conti001}) 
and~(\ref{eqn:tqunitmass}), one is led to the following radial diffusion equation:
{\setlength\arraycolsep{2pt}
\begin{eqnarray}
\label{eqn:conti1}
\frac{\partial\Sigma}{\partial t}&=&
\frac{1}{2\pi}\left(\frac{r'}{r}|\Omega_p\right.
-\Omega'|\Sigma'
\frac{\delta r'}{\delta r}{}
{}-|\Omega_p-\Omega|\Sigma'\left)\Pi\left(\frac{|x|}{x_s}\right)
+3\nu\frac{\partial^2\Sigma}{\partial r^2}\right. -\frac{1}{\pi r^2\Omega}
\frac{d^2\Gamma}{dr^2}
\end{eqnarray}}
where $\Pi$ is defined as:
\begin{displaymath}
\begin{array}{ccl}
\Pi:u &\longmapsto& 1\mbox{~~if $|u|<1$}\\
& & 0\mbox{~~otherwise}
\end{array}
\end{displaymath}
and accounts for the fact that the source and sink terms are present
only over the horseshoe region (i.e. $|x|<x_s$).
In Eq.~(\ref{eqn:conti1})  we have assumed
$r|\partial^2\Sigma/\partial r^2| \gg |\partial\Sigma/\partial r|$ and
$r|d^2\Gamma/dr^2|\gg|d\Gamma/dr|$.
If we neglect both pressure and viscous effects during a
U-turn at one end of the horseshoe orbit, then the
conservation of the Jacobi constant of a fluid element implies, as has
been shown by Ward (1992), that:
\begin{equation}
\label{eqn:conti2}
\frac{\delta r}{\delta r'}=
\frac{r'B'|\Omega'-\Omega_p|}{rB|\Omega-\Omega_p|}
\end{equation}
where $B$, the second Oort's constant, in the case of a Keplerian
 disk,
 is also $B=\Omega/4$.
Using Eq.~(\ref{eqn:conti1}) and~(\ref{eqn:conti2}), one can write:
{\setlength\arraycolsep{2pt}
\begin{eqnarray}
\label{eqn:conti3}
\frac{\partial\Sigma}{\partial t}&=&
\frac{1}{2\pi}\left(\Sigma'\frac{\Omega}{\Omega'}-\Sigma\right)
|\Omega_p-\Omega|\Pi\left(\frac{|x|}{x_s}\right)
+3\nu\frac{\partial^2\Sigma}{\partial r^2}
-\frac{1}{\pi r_p^2\Omega_p}\frac{d^2\Gamma}{dr^2}
\end{eqnarray}}
If we write $x'=r'-r_p$, and if we define the dimensionless distances
to the orbit as $\hat x=x/r_p$ and $\hat x'=x'/r_p$, then we have:
\begin{equation}
\frac{\Omega}{\Omega'}=1-\frac 32\hat x+\frac 32\hat
x'+\frac{15}{8}\hat x^2+\frac 38\hat x'^2+O(\hat x^3)+O(\hat x'^3)
\end{equation}
From Eq.~(\ref{eqn:conti2}), we can infer that $\hat x=-\hat x'$ to lowest
order in $\hat x$, therefore:
\begin{equation}
\hat x'=-\hat x+\zeta \hat x^2+O(\hat x^3)
\end{equation}
where $\zeta$ is a numerical constant. Hence the
bracket in the source term of Eq.~(\ref{eqn:conti3}) reads:
\begin{eqnarray}
\Sigma'\frac{\Omega}{\Omega'}-\Sigma&=&\Sigma(-\hat x)-\Sigma(\hat
x)-3\hat x\Sigma(-\hat x){}\\\nonumber
& &{}+\frac 14(9+6\zeta)\hat x^2\Sigma(-\hat x)+
\zeta \hat x^2\frac{\partial\Sigma}{\partial \hat x}(-\hat x)+O(\hat x^3)
\end{eqnarray}
If we keep only terms to first order in $\hat x$ in this expression,
Eq.~(\ref{eqn:conti3})
can be recast as:
\begin{eqnarray}
\label{eqn:diff1}
\frac{\partial\Sigma}{\partial t}&=&\frac{3}{4\pi}\Omega_p[\Sigma(-\hat x)-\Sigma(\hat
x)-3\hat x\Sigma(-\hat x)]{}\\\nonumber
& & {}\times\left(1-\frac 54 \hat x\right)|\hat
x|\Pi\left(\frac{\hat x}{\hat x_s}\right)+\frac{3\nu}{r_p^2}\frac{\partial^2\Sigma}{\partial \hat x^2}-\frac{1}{\pi r_p^4\Omega_p}\frac{d^2\Gamma}{d\hat x^2}+O(\hat x^3)
\end{eqnarray}
We call $\tilde\Sigma$ and $\check\Sigma$ respectively the odd
and even part in $\hat x$ of $\Sigma$:
\begin{displaymath}
\left\{\begin{array}{rcl}
\tilde\Sigma(\hat x)&=&\frac 12[\Sigma(\hat x)-\Sigma(-\hat x)]\\
\check\Sigma(\hat x)&=&\frac 12[\Sigma(\hat x)+\Sigma(-\hat x)]
\end{array}\right.
\end{displaymath}
We write $\check\Sigma(\hat x)=\Sigma_c+O(\hat x^2)$, where
$\Sigma_c=\check\Sigma(0)
=\Sigma(0)$.
Taking the odd part in $\hat x$ of both sides of~Eq.~(\ref{eqn:diff1}), and keeping
terms only up to $O(\hat x^2)$,
one is led to:
\begin{eqnarray}
\frac{\partial\tilde\Sigma}{\partial t}&=&-\frac{3}{4\pi}\Omega_p|\hat
x|\Pi\left(\frac{\hat x}{\hat x_s}\right)[2\tilde\Sigma(\hat x)+3\hat
x\Sigma_c]
+\frac{3\nu}{r_p^2}\frac{\partial^2\tilde\Sigma}{\partial\hat x^2}
\end{eqnarray}
where we have neglected, at this step, the Inner and Outer Lindblad torque imbalance,
and thus the odd part in $\hat x$ of $\Gamma$ and its second derivative vanishes.
We write $s=\tilde\Sigma/\Sigma_c$ and
$\hat\nu=\nu/(r_p^2\Omega_p)$, and we assume that we have a steady
state. Therefore, for $\hat x>0$, we have:
{\setlength\arraycolsep{2pt}
\begin{eqnarray}
\label{eqn:diffset}
\hat\nu\frac{\partial^2s}{\partial\hat
x^2}&=&\frac{1}{4\pi}[2s+3\hat x]\hat x\mbox{~~if $\hat x<\hat x_s$}\\ \nonumber
&=&{}0\mbox{~~otherwise}
\end{eqnarray}
}
The first line of Eq.~(\ref{eqn:diffset}) shows the balance between
libration, which tends to create a profile of density such that
$s=-\frac 32\hat x$, and the viscous diffusion. We want to find a
solution to this ODE valid for any $\hat x>0$, and such that $s(0)
=0$, since $s$ is an odd function of~$x$.

The general solution of~Eq.~(\ref{eqn:diffset}) for $\hat
x>\hat x_s$ is $s=a\hat x+b\hat x^2$, which does not diverge for
large values of $\hat x$  only if $a=0$
and $b=0$. Therefore, we are led to seek a solution of
Eq.~(\ref{eqn:diffset})
for $\hat x\in[0,\hat x_s]$ which fulfills the two conditions:
\begin{equation}
\label{eqn:condi1}
\left\{\begin{array}{rcl}
s(0)&=&0\\
s'(\hat x_s)&=&0
\end{array}\right.
\end{equation}
where $s'\equiv\partial s/\partial\hat
x$.
If we define 
\begin{equation}
\label{eqn:defs}
S=s+\frac 32\hat x
\end{equation}
and 
\begin{equation}
\label{eqn:defz}
z=(2\pi\hat\nu)^{-1/3}\hat x,
\end{equation}
 then
Eq.~(\ref{eqn:diffset})
can be recast, for $0<z<z_s=(2\pi\hat\nu)^{-1/3}\hat x_s$, as:
\begin{eqnarray}
\label{eqn:diff2}
\frac{\partial^2S}{\partial z^2}=zS
\end{eqnarray}
and the set of conditions~(\ref{eqn:condi1}) becomes:
\begin{equation}
\label{eqn:condi2}
\left\{\begin{array}{rcl}
S(0)&=&0\\
\frac{\partial S}{\partial z}(z_s)&=&\frac 32(2\pi\hat\nu)^{\frac 13}
\end{array}\right.
\end{equation}
The solution of Eq.~(\ref{eqn:diff2}) which fulfills the
conditions~(\ref{eqn:condi2})
is:
\begin{equation}
\label{eqn:sol1}
S(z)=\frac{3(2\pi\hat\nu)^{1/3}}{2g'(z_s)}g(z)
\end{equation}
where:
\begin{equation}
\label{eqn:defg}
g(z)=\frac{\mbox{Bi}(z)-\sqrt{3}\mbox{Ai}(z)}{2\mbox{Bi}'(0)}
\end{equation}
is the linear combination of Airy functions Ai and Bi such that $g(0)=0$ and
$g'(0)=1$.
This solution can be used to evaluate the coorbital corotation torque
given by the expressions~(\ref{eqn:ct1}), (\ref{eqn:vtp}) and 
(\ref{eqn:gammar2}). In this part we neglect the perturbed azimuthal
velocity, and we neglect as well the even part of the density profile
perturbation
(the shallow `dip') which appears around the orbit for a sufficient
mass (i.e., we write $\check\Sigma\equiv\Sigma_\infty$). 
Furthermore we assume that the mass flow across the coorbital region from
the outer disk to the inner disk is $\dot M=3\pi\nu\Sigma_\infty$.
 Therefore the expression of the corotation torque, in this
approximation, reduces to:
\begin{eqnarray}
\label{eqn:ctsimpl1}
\Gamma_C&=&3\pi\nu[(r_p-x_s)^2\Omega(r_p+x_s)^2\Sigma_{S^+}{}\\ \nonumber
& & {}-(r_p-x_s)^2\Omega(r_p-x_s)\Sigma_{S^-}]\\ \nonumber
& & {}+3\pi\nu\Sigma_\infty[j_0(r_p+x_s)-j_0(r_p-x_s)]
\end{eqnarray}
which leads to:
\begin{equation}
\label{eqn:ctsimpl2}
\Gamma_C=-6\pi\nu r_p^2\Omega_p\Sigma_\infty s(\hat x_s)
\end{equation}
Using Eqs.~(\ref{eqn:defs}),  (\ref{eqn:sol1}) and~(\ref{eqn:ctsimpl2}), one can finally
deduce~Eq.~(\ref{eqn:gammac}).

\section{Additional terms}
\label{otherterms}
The additional term arising from the even part in $\hat x$ of 
$\partial v/\partial r$ in
Eq.~(\ref{eqn:vtp})
reads:
\begin{equation}
\label{eqn:term2}
\Gamma_C^{II}=4\pi\nu r_px_s\Sigma_c\left(\frac{\partial v}{\partial
r}\right)_{S+}(2-\beta)
\end{equation}
where $\Sigma_c$ is the surface density at the planet orbit.

In order to give an estimate of $v$, we write the axisymmetric part of
the linearized radial Euler equation, in which we omit the gradient
of the $m=0$ perturbed potential of the planet. This reads:
\begin{equation}
\label{eqn:lim1}
2\Omega v-c_s^2\frac{\partial_r\Sigma}{\Sigma_\infty}=0
\end{equation}
where $c_s$ is the sound speed.
We can give a maximal estimate of $\partial_r\Sigma$ by writing
that the wave damping occurs over a radial range of at least one disk
thickness.
This reads:
\begin{equation}
\label{eqn:lim2}
3\pi\nu\Omega r^2\frac{\partial\Sigma}{\partial
r}\le\frac{\Gamma_{LR}}{h}
\end{equation}
Using Eqs.~(\ref{eqn:lim1}) and~(\ref{eqn:lim2}) one infers:
\begin{equation}
\label{eqn:lim3}
v\le\frac{h}{6\pi\nu\Sigma_\infty r^2}\Gamma_{LR}
\end{equation}
In the same manner one can write an upper limit for the perturbed
azimuthal velocity derivative: $\partial_rv\le v/h$. Therefore:
\begin{equation}
\label{eqn:lim4}
\frac{\partial v}{\partial r} \le
\frac{\Gamma_{LR}}{6\pi\nu\Sigma_\infty r^2}
\end{equation}
Assuming that $\Sigma_c\simeq\Sigma_\infty$, i.e., the
dip surrounding the planet orbit is shallow, one can infer 
from Eqs.~(\ref{eqn:term2}) and~(\ref{eqn:lim4}) that:
\begin{equation}
\Gamma_C^{II} \le \frac 23(2-\beta)\frac{x_s}{r_p}\Gamma_{LR}
\end{equation}
The additional term arising from~$v$ in Eq.~(\ref{eqn:gammar1}) reads:
\begin{equation}
\Gamma_C^{III} = 3\pi\nu\Sigma_\infty\cdot 2x_sv
\end{equation}
Using Eq.~(\ref{eqn:lim3}), one finds as a maximal value for this
term:
\begin{equation}
\Gamma_C^{III} \le \frac{x_sh}{r_p^2}\Gamma_{LR}
\end{equation}
Since $h\ll r_p$ and $x_s\ll r_p$, this term is negligible.

\begin{acknowledgements}
I wish to thank Prof. J.C.B. Papaloizou for discussions and advice, and
an anonymous referee whose comments greatly enhanced the first version
of this manuscript.
This work was supported by the research
network ``Accretion onto black holes, compact stars and protostars''
funded by the European Commission under contract number ERBFMRX-CT98-0195.
\end{acknowledgements}

\end{document}